\documentclass[prb,twocolumn,showpacs]{revtex4-1}
\usepackage{graphicx}

\begin{document}

\title{The spectroscopic signature of Kondo screening on single adatoms in Na(Fe$_{0.96}$Co$_{0.03}$Mn$_{0.01}$)As}

\author{Zhenyu Wang$^{1}$, Delong Fang$^2$, Qiang Deng$^2$, Huan Yang$^{2,*}$, Cong Ren$^1$ and Hai-Hu Wen$^{2,*}$}

\affiliation{$^1$National Laboratory for Superconductivity, Institute of Physics and National Laboratory for Condensed Matter Physics, Chinese Academy of Sciences, Beijing 100190, China}

\affiliation{$^2$Center for Superconducting Physics and Materials, National Laboratory of Solid State Microstructures and Department of Physics, Nanjing University, Nanjing 210093, China}

\begin{abstract}
The electronic states of surface adatoms in Na(Fe$_{0.96}$Co$_{0.03}$Mn$_{0.01}$)As have been studied by low temperature scanning tunneling
spectroscopy. The spectra recorded on the adatoms
display both superconducting coherence peaks and an asymmetric resonance
in a larger energy scale. The Fano-type line shape of the spectra
points towards a possible
Kondo effect at play. The apparent energy position of the resonance peak shifts about $5\;$meV to the Fermi level when measured across the
critical temperature, supporting that the Bogoliubov quasiparticle is responsible for the Kondo screening in the superconducting state. The tunneling spectra do not show
the subgap bound states,
which is explained as the weak pair breaking effect given by the weak and broad scattering potential after the Kondo screening.
\end{abstract}


\maketitle

According to the Bardeen-Cooper-Schrieffer theory, metals become superconducting when Cooper pairs are
formed via electron-phonon interaction and condense below the critical temperature ($T_\mathrm{c}$).
The Cooper pairs with the $s$-wave symmetry can survive in the presence of non-magnetic impurities,
while magnetic impurities are expected to suppress superconductivity and induce bound states within the superconducting gap\cite{Anderson,BalatskyRevModPhys}. The
energy of the in-gap state is dominated by the interaction strength between the magnetic impurities and the Cooper pairs\cite{FlattePRL}.
On the other hand, in normal metals containing magnetic impurities, the magnetic coupling
between the local impurity spin and the itinerant electrons can lead to another singlet formation, namely to
form a Kondo screening cloud, at temperature below the characteristic Kondo temperature ($T_\mathrm{K}$) of the system\cite{Kondo,Hewson}.
The competition between the Kondo screening and the opening of superconducting gap can be described by the ratio of $k_\mathrm{B}T_\mathrm{K}$ to $\Delta$\cite{SatoriJPSJ},
where $k_\mathrm{B}$ is the Boltzmann constant and the order parameter $\Delta$ governs the superconducting pairing strength.

Low temperature scanning tunneling microscopy (STM) and spectroscopy (STS) provide a direct method to study the local density of states (LDOS) near the impurities.
The in-gap states induced by magnetic impurities (Mn, Gd)
adsorbed on the surface of Nb single crystal have been proved\cite{YazdaniScience}, in consistent with the pioneer calculation of Yu-Shiba-Rusinov\cite{Yu,Shiba,Rusinov}. In the
high temperature superconductors, besides magnetic impurities, non-magnetic Zn substitution in  Bi$_2$Sr$_2$CaCu$_2$O$_{8+x}$ and Cu substitution in the optimally doped Na(Fe$_{1-x}$Co$_{x}$)As also showed clear in-gap resonant states, providing strong evidence of the sign-reversal order parameters in those materials\cite{PanZnImpurity,HyangNC}.
Meanwhile, STM experiments have probed various kinds of magnetic atoms absorbed on metallic surfaces, and the Kondo resonance-like structures have also been
observed in the differential conductance near the Fermi energy\cite{MadhanvanScience,JamnealaPRB}. The asymmetric line shapes resemble that
of a Fano resonance\cite{Fano}. Moreover, in a system with variable localized magnetic coupling strengths, a balance
between Kondo screening and superconducting pair-breaking interaction has been revealed\cite{FrankeScience}. The result showed that the energy
of the bound state is located close to the superconducting gap edge for $k_\mathrm{B}T_\mathrm{K} \gg \Delta$ and deeply in the gap for $k_\mathrm{B}T_\mathrm{K} \sim \Delta$.
In iron pnictides, the superconductivity is widely believed to have a magnetic origin\cite{Chubukov}. Therefore it deserves further investigation to identify the interplay of superconducting pairing and the Kondo screening when a magnetic impurity is presented in those compounds.

\begin{figure}
\includegraphics[width=8cm]{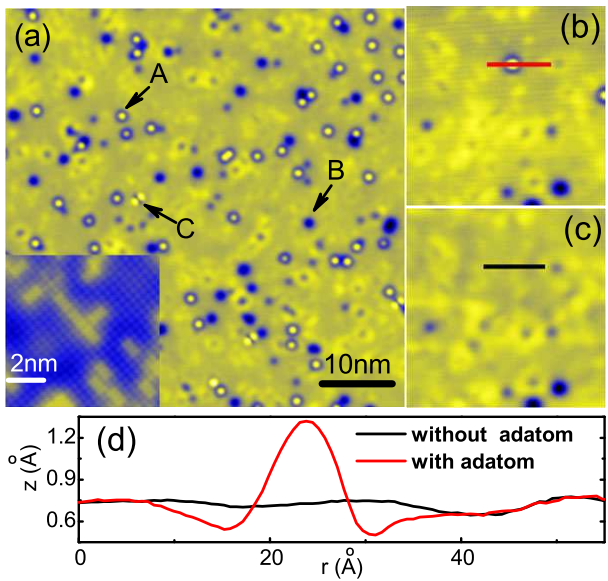}
\caption{(Color online) (a) Topographic STM
image of the Na(Fe$_{0.96}$Co$_{0.03}$Mn$_{0.01}$)As with three kinds of typical defects.
Inset: Atomic-resolution topography in a defect free region.
Each of the $2\times1$ rectangular block corresponds to a Co-impurity.
(b,c) Topographies in a $17.5$ nm $\times 17.5$ nm region before and
after the ``picking-up'' process, respectively (see text). (d) The spatial dependence
of height $z$ measured along the red and dark lines marked in (b) and (c).
The tunneling conditions were $V_\mathrm{s}$ = 130 mV and $I_\mathrm{t}$ = 50 pA for (a), (b) and (c),
$V_\mathrm{s}$ = 50 mV and $I_\mathrm{t}$ = 100 pA for the inset of (a). } \label{fig1}
\end{figure}

High quality Na(Fe$_{0.96}$Co$_{0.03}$Mn$_{0.01}$)As single crystals ($T_\mathrm{c} \approx 12.8\;$K)
were synthesized by the flux method\cite{QDengsample}.
All STM and STS tunneling measurements were carried out with an ultrahigh-vacuum
low-temperature scanning probe microscope USM-1300 (Unisoku). The samples were cleaved in an
ultra-high vacuum with a base pressure about $1.1 \times 10^{-10}\;$Torr,
then immediately inserted into the microscopy head, which was kept at a low temperature. In all
measurements, Pt/Ir tips were used. The tunneling spectra were acquired using the lock-in
technique with ac modulation of $0.8\;$mV at $987.5\;$Hz.

\begin{figure}
\includegraphics[width=8cm]{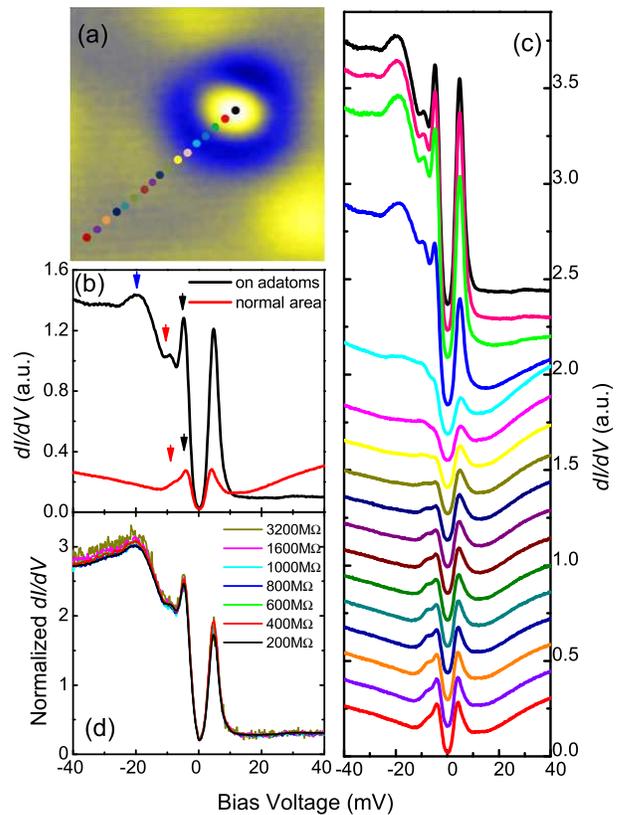}
\caption{(Color online) (a) Zoom in on a 4.3 nm $\times$ 4.3 nm area containing an adatom, which was taken
with $V_\mathrm{s}$ = 100 mV and $I_\mathrm{t}$ = 50 pA.
(b) Typical tunneling spectra measured on and off the adatom in (a) at $1.7$ K.
(c) Spatially resolved spectra of $dI/dV$ versus bias voltage recorded along the trajectory ($17$ points) indicated in (a).
The spectra have been shifted vertically for clarity.
(d) Tunneling spectra with various tunneling current set-points. The data
have been normalized to the conductance value at $V_\mathrm{s} = 40$ meV.} \label{fig2}
\end{figure}

Fig.~\ref{fig1}(a) depicts a constant current topographic image with three
types of defects on the cleaved surface of Na(Fe$_{0.96}$Co$_{0.03}$Mn$_{0.01}$)As: round protrusions
(labeled A), vacancy-like depressions (labeled B) and orthogonal dumbbell-like features (labeled C).
The inset shows an atomic-resolution image we achieved far away from those defects, with atomic
spacing $\sim 3.9\;${\AA}. The weakest bond in this crystal sits between the two
adjacent Na layers, therefore the topmost layer after cleaving is the one with Na atoms.
Beside the observed square lattice, one can see some rectangular blocks, which have already been
identified as the Co dopants on the Fe-layer\cite{YangPRB}. For the Mn substitutions, it can only be identified with lower bias voltage during the scanning.
The type B defects, which may correspond to the missing atoms, were widely distributed
on the ever-scanned surface, and no significant spectroscopic differences were detected spatially. For the type C defects, the electronic behavior needs further study due to the very low coverage.
The coverage of type A defects varied dramatically in different places. In most area,
the type A defects were absent. Intriguingly, this kind of defects can be removed
by bringing the STM tip close enough to the surface. For example, Fig.~\ref{fig1}(b) and (c) present
images of the same area before and after making such action.
We therefore argue that type A defects
are adatoms sitting on the surface. Fig.~\ref{fig1}(d) displays line profiles, in terms of height variation,
obtained along the red line marked in Fig.~\ref{fig1}(b) and the black line in Fig.~\ref{fig1}(c), respectively.
The apparent height of the adatom is about $0.7\;${\AA}, measured as the distance between peak and the average
surrounding background level. We focus the rest of this paper on the electronic properties near these adatoms.

In Fig.~\ref{fig2}(b) the tunneling spectra recorded at 1.7 K on and off a single adatom are reported. The red
solid curve shows $dI/dV$ spectrum measured with the tip held over defect-free surface.
The spectrum exhibits two clear peaks at an energy level of about $\pm4.2\;$meV (marked with black arrow in the negative bias side), which are associated
with the superconducting gap.
An additional conductance feature appears at higher energy (marked with red arrow), reminiscent of the
bosonic mode observed in the Na(Fe$_{0.975}$Co$_{0.025}$)As single crystal\cite{WangNP}.
In the sample we studied here, the mode energy of $\Omega \sim 5\;$meV $\sim 4.5k_\mathrm{B}T_\mathrm{c}$, closely agrees with our previous work.
The black solid curve displays spectrum recorded with the tip held over the center of an
adatom.
The superconducting coherence peaks and the bosonic mode structures are clearly
visible, or even get enhanced. Additionally, the spectrum shows a pronounced background peak well outside the superconducting gap (marked with blue arrow), and
the LDOS of the unoccupied side is strongly suppressed. These features were observed on hundreds of adatoms at different locations,
and the peak position varies with location, ranging from $-14$ to $-25\;$meV.
This kind of adatoms has also been detected on the surface of Na(Fe$_{0.975}$Co$_{0.025}$)As with similar electronic behavior.
In Fig.~\ref{fig2}(c), we display a set of tunneling spectra with tip held at varying distance from the center of an adatom indicated in Fig.~\ref{fig2}(a).
The resonance-like background feature both decreases in amplitude and changes shape as measured outward. At the position of
about $15\;${\AA} from the center of the adatom, the feature is mostly gone.
We also measured a series of $dI/dV$ spectra at various tunneling current set-points, i.e., with junction resistances ranging from $200$ to $3200\;$M$\Omega$.
Fig.~\ref{fig2}(d) shows the spectra which have been normalized to the conductance value at $40\;$meV. No significant change was observed over this resistance range, which
can rule out some tip induced effects\cite{TIBB}.

\begin{figure}
\includegraphics[width=8cm]{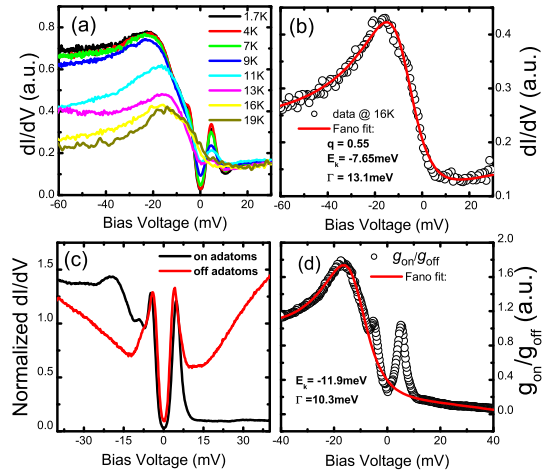}
\caption{(Color online) (a) The evolution of the tunneling conductance at an adatom site with temperature
increased from 1.7 K to 19 K. (b) Spectrum taken at $16$ K, when the sample is in normal state. The full line
is a Fano fit to the curve, and the parameter of the fit is given in the figure. (c) Typical tunneling differential conductance ($g$) measured on and off the adatom,
which have been normalized to the value at about $-5$ mV. (d) $g_\mathrm{on}/g_\mathrm{off}$, which was successfully
fitted with the modified Fano function.} \label{fig3}
\end{figure}

Fig.~\ref{fig3}(a) shows the evolution of the STS spectra with
temperature up to $19\;$K, and it is clear that the pronounced asymmetric background lineshape
tends to become more symmetric and the resonance peak moves slightly to the Fermi energy as the temperature is increased. Above $13\;$K, when the sample is in the normal
state, the superconducting coherence peaks disappear and the curves show only the asymmetric resonance feature. This type of asymmetric spectra with dip or peak features near the Fermi energy have been observed
for the individual magnetic impurity on metal surface and interpreted as the Kondo resonance\cite{MadhanvanScience,JamnealaPRB,LiPRL}.
Below $T_\mathrm{K}$, the spin of impurity can be flipped by an itinerant electron while simultaneously a spin excitation state called the Kondo resonance is created close to
the Fermi energy. This spin exchange can modify the energy spectrum of the system. In STS measurement, an electron tunneling from
a tip to the Kondo resonance actually has two different paths with different probabilities (the orbital of impurity and the continuum), leading
to a quantum interference term. This term gives rise to a so-called Fano lineshape and described as
\begin{equation}\label{equ.1}
\rho(E) \propto \rho_\mathrm{0} + \frac{(q+\epsilon)^2}{1+\epsilon^2}
\end{equation}
where $\epsilon = (E - E_\mathrm{K})/\Gamma$ is the normalized energy, with $E_\mathrm{K}$ as the energy position of the resonance from the Fermi level and
$\Gamma = k_\mathrm{B}T_\mathrm{K}$ as the half-width at half-maximum of the curve. The parameter $q$ is related to
the interference of the two channels contributing to the Fano line shape\cite{Ujsaghy}, and the Fano resonance shape thus depends on $q$.
Fig.~\ref{fig3}(b) shows the spectrum obtained at $16\;$K and
the corresponding fit to Eq.~\ref{equ.1}, which yields a half-width of $\Gamma \sim 13.1\;$meV.
To investigate how these adatoms modify the spectra in superconducting state, we normalized the spectra obtained on and off the adatoms at $1.7\;$K
at energy of about $-5\;$meV, and then used Fano equation to fit $g_{on}/g_{off}$ ($g = dI/dV \sim$ LDOS). The outcomes are depicted in Fig.~\ref{fig3}(c) and Fig.~\ref{fig3}(d),
respectively. One excellent fit of the spectrum outside the superconducting feature is achieved by introducing a linear background term into Eq.~\ref{equ.1}\cite{BorkNP}, and yields a
value of $k_\mathrm{B}T_\mathrm{K} \sim 10.5\;$meV and thus $T_\mathrm{K}=122\;$K.
With the excellent agreement to the Fano fitting of the tunneling spectra, we argue that the adatoms are most likely to
be excess magnetic impurities, such as Fe or Co, which induce a strong Kondo resonance in the vicinity.

\begin{figure}
\includegraphics[width=8cm]{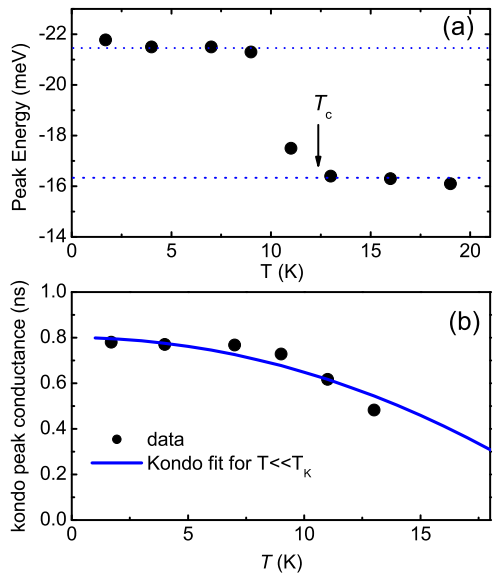}
\caption{(Color online) (a) Temperature dependence of the resonance peak energy directly extracted from Fig.~\ref{fig3}(a).
The absolute value decreases from $21.5$meV to $16.4$meV. The blue short dot lines are to guide the eye. (b) Temperature dependence of the magnitude of resonance peak height,
which can be understood with the Kondo effect model (blue line).} \label{fig4}
\end{figure}

Fig.~\ref{fig3}(a) allows us to semi-quantitatively determine the temperature dependence of the resonance characters. A close look into the variation of the resonance
peak position reveals that it moves to Fermi energy slightly with elevating temperature. As a result, Fig.~\ref{fig4}(a) shows the temperature dependence of the resonance peak energy
directly extracted from the $dI/dV$ curves. The absolute value decreases from $21.5\;$meV to $16.4\;$meV, which cannot be simply explained by the thermal broadening.
Interestingly, the resonance peak almost does not change its location at temperature below $9\;$K or above $T_\mathrm{c}$.
This decrease can be rationalized qualitatively considering that the Bogoliubov quasiparticle is responsible for
screening the local spin of the adatom, which needs an energy of $\Delta(T)$ to overcome the superconducting pairing potential. Furthermore, the resonance peak energy changes
slowly at low temperature but drop down very quickly near $T_\mathrm{c}$, giving further support to this picture. Similar temperature dependent behavior were observed on other adatoms
in our measurements.

Another intriguing observation is that, with elevating temperature,
the asymmetric background of the lineshape is gradually washed out. One important characteristics of the Kondo effect is the broadening and reduction of the Kondo resonance
with increasing temperature\cite{HankePRB}. In the Fermi liquid model, the Kondo peak height has a temperature dependent behavior and is predicted to decay
slowly with $1 - c(T/T_\mathrm{K})^2$ for $T \ll T_\mathrm{K}$\cite{Nozieres}. However, determining the absolute intensity of the conductance for different spectra in STS measurements especially with
varying temperature is a challenge.
Considering that the LDOS at the positive bias side is relatively featureless, we used a simplest analytical treatment to illustrate the temperature dependence of the resonance feature, i.e., we normalized the spectra to the conductance value at about $20\;$meV and then extracted the magnitude of the peak in the normalized curves as a description of the Kondo resonance peak height. The result arising from this process and a theoretical fit with $c/T_\mathrm{K}^2=0.0020\pm0.0002$ are shown in Fig.~\ref{fig4}(b). One can see that the decrease of the peak height follows the expected behavior of a Kondo resonance at low temperature.

No clear in-gap bound states were detected on these magnetic adatoms, in striking contrast with the pervious STS measurement on Mn impurities in this material\cite{HyangNC} and the conventional superconductor Nb\cite{YazdaniScience}. This seems to create a dilemma since the magnetic impurities are always regarded as pair-breaker in superconductors.
Nevertheless, the competition between Kondo screening and superconductivity should be taken into account. The absence of in-gap states was also detected for
adsorbed Co adatom on the Cu (111) surface using a superconducting
Nb tip\cite{Ternes}.
STS performed on manganese-phthalocyanine molecules on Pb(111) substrate revealed the evolution from a Kondo screened singlet ground state to a unscreened multiplet ground state,
and the in-gap states was found to be induced very close to the gap edge when $k_\mathrm{B}T_\mathrm{K}/\Delta \gg 1$\cite{FrankeScience}. The absence of the in-gap bound states in our present case may be understood as a consequence of the $s\pm$ pairing together with the Kondo screening effect on the spin of the adatom. In iron pnictides, superconducting pairing is suggested to be established by exchanging the spin fluctuations leading to a gap with reversed signs on the hole and electron pockets\cite{Mazin,Kuroki,Lee,Hirschfeld}.
STM measurements involving impurity effects have been successfully performed and many interesting features have been reported\cite{Hoffman,Davis,Song,Pennec,Hanaguri,Golden}.
For the adatom we studied here, the larger energy scale of the Kondo resonance channel, with
$k_\mathrm{B}T_\mathrm{K} \sim 3\Delta$, may give rise to the strong screening of the local spin of the adatom. At the mean time, the scattering potential gets much broader due to
this Kondo screening. For the inter-pocket $s\pm$ pairing with large momentum transfer during the pair-scattering process, a broadened scattering potential can hardly
act as the effective pair breaker. Both effects mentioned here will weaken the pair-breaking.
This is quite similar to the case of Co dopants in Na(Fe$_{1-x}$Co$_{x}$)As\cite{YangPRB}. Therefore, although the adatoms discovered in our experiment can produce clear
Kondo effect and modify the tunneling spectrum significantly, the absence of in-gap bound states may however gain an explanation showing the consistency with the $s\pm$ pairing.

In summary, we have presented spatial evolution and temperature dependence of the tunneling spectra associated with the adatoms on
iron pnictide Na(Fe$_{0.96}$Co$_{0.03}$Mn$_{0.01}$)As. At the adatom and below $T_c$, the superconducting spectrum is significantly modified with a Kondo resonance like background. The temperature dependence of the resonance is in good agreement with the predicted Kondo behavior, giving further evidence that the Kondo
effect is playing the role here. The spectra in the superconducting state reveal the absence of in-gap bound states predicted theoretically
for magnetic impurities. This is understood as the consequence of the Kondo screening effect on the spin of the adatom, and the broadened scattering potential, both will weaken the pair breaking effect with the scenario of $s\pm$ pairing.

This work is supported by the Ministry of Science and Technology
of China (973 projects: 2011CBA00102, 2012CB821403,
2010CB923002), NSF and PAPD of China.

$^*$ huanyang@nju.edu.cn, hhwen@nju.edu.cn

\end{document}